\documentclass[runningheads]{llncs}
\usepackage{graphicx} 
\usepackage{appendix}
\usepackage{amsmath}
\usepackage{amsfonts}
\usepackage{amssymb}
\usepackage{amsmath,mathtools}
\usepackage{mathtools}
\usepackage{paralist}
\usepackage{xcolor}
\usepackage{setspace} 
\usepackage{hyperref}
\usepackage{todonotes}
\usepackage{csquotes,comment}

\usepackage[utf8]{inputenc} 
\usepackage[T1]{fontenc}

\usepackage{amsmath}
\usepackage{amsfonts}
\usepackage{amssymb}

\usepackage[adversary,operators,mm,primitives,keys,complexity,advantage,notions]{cryptocode}

\usepackage{caption}
\usepackage{cite}
\usepackage{url}

\newcommand\pesto{\texttt{Pesto}}

\newcommand\Fq{\mathbb{F}_{q}}
\newcommand\Fqn{\mathbb{F}_{q}^n}
\newcommand\Fqm{\mathbb{F}_{q}^m}

\def\q{{\mathfrak q}}
\def\fq{{\mathbb F}_{q}}
\newcommand{\ideal}[1]{\left({#1}\right)}
\newcommand{\vs}[1]{\langle{#1}\rangle}

\title{Cryptanalysis of a multivariate CCZ scheme}
\author{Alessio Caminata\inst{1}  \and Elisa Gorla\inst{2}  \and  Madison Mabe\inst{3}  \and Martina Vigorito\inst{4} \and Irene Villa\inst{1,5} }

 \institute{Universit\`a di Genova
 \and
Universit\'e de Neuch\^atel
 \and
Clemson University 
 \and Universit\`a di Salerno
 \and Universit\`a di Trento 
 }

  \authorrunning{A. Caminata et al.}

\date{}

\begin{document}

\maketitle

\begin{abstract}
We consider the multivariate scheme \texttt{Pesto}, which was introduced by Calderini, Caminata, and Villa. In this scheme, the public polynomials are obtained by applying a CCZ transformation to a set of quadratic secret polynomials. As a consequence, the public key consists of polynomials of degree $4$. 
In this work, we show that the public degree $4$ polynomial system can be efficiently reduced to a system of quadratic polynomials. This seems to suggest that the CCZ transformation may not offer a significant increase in security, contrary to what was initially believed.
\end{abstract}

\section{Introduction}

Multivariate cryptography is one of the main candidates for providing quantum resistant public key primitives.
The security of multivariate cryptography is based on the Polynomial System Solving Problem (PoSSo Problem), that is, solving a random system of multivariate polynomials of degree at least $2$. This problem is known to be NP-complete and believed to be hard in the average case, already for quadratic polynomials over the field $\mathbb{F}_2$.

In the usual setup of multivariate schemes, there is a central (secret) map $F$ consisting of $m$ polynomials in $n$ variables. These polynomials (usually quadratic) are not random, but rather possess a trapdoor that allows the legitimate user to easily find a solution to the system. The map $F$ is then composed with two randomly chosen affine linear bijections, $A_1$ and $A_2$, respectively over $\mathbb{F}_q^m$ and $\mathbb{F}_q^n$, to obtain the public map $G_{\mathrm{pub}} = A_1 \circ F \circ A_2$. The public map consists of polynomials that resemble random polynomials and is therefore assumed to be hard to invert. The secret and public keys, $F$ and $G_{\mathrm{pub}}$, are called \textit{affine equivalent}.

Despite the use of affine random bijections, the maps $F$ and $G_{\mathrm{pub}}$ often share algebraic properties, that have been exploited in the past to attack several proposed cryptographic primitives. In an attempt to propose an alternative construction, Calderini, Caminata, and Villa~\cite{CCV24} suggested applying more general transformations  to the central map, in order to obtain a public-key function $G_{\mathrm{pub}}$ that does not directly inherit the \enquote{simple} structure of the secret function $F$. Specifically, they propose using the CCZ transformation, introduced by Carlet, Charpin, and Zinoviev~\cite{CCZ} in the context of Boolean functions.
One potential advantage of the CCZ transformation is that it usually produces public polynomials of higher degree than the secret polynomials, making them potentially more difficult to solve, while keeping the secret key relatively small.
The authors of~\cite{CCV24} presented a general instance of the scheme, which can be applied to any central map admitting a $t$-twist, see~\cite{CP19}. Then, they presented some restrictions on the choice of the quadratic secret map used, summarized in a concrete instance with a specific choice for the central map (see~\cite[\S4.2.3]{CCV24}), which they call \texttt{Pesto}, on which we will focus throughout this paper.

In the  \texttt{Pesto} scheme, the central map consists of quadratic equations in $n$ variables that, once composed with the CCZ transformation, contains equations of degree up to $4$. Thus, the public key $G_{\mathrm{pub}}$ is of degree 4, while the secret key $F$ is of degree $2$. In this paper, we show that the equations of the public key can be efficiently reduced to equations of degree $2$. This suggests that there is no clear advantage in terms of security in composing with the CCZ transformation.

In this paper, we discuss two techniques for carrying on this reduction: Gr\"obner bases algorithms and Higher Order Linearization Equations. The second method seems to be asymptotically faster, with a complexity of $\mathcal{O}(n^{7})$, when $m\leq n$. Both methods take advantage of the structure of the central map for the reduction. Once $G_{\mathrm{pub}}$ is reduced to a quadratic polynomial system, it can be solved by using general quadratic solving methods or any OV solving methods, if it has the structure of an Oil and Vinegar (OV) scheme.

\subsubsection{Acknowledgments}

This work started during the VT--Swiss Coding Theory \& Cryptography Summer School, held in Riva San Vitale in July 2024. We thank the organizers for providing us with a productive working environment.\\
A.~Caminata is supported by the Italian PRIN2022 grants P2022J4HRR ``Mathematical Primitives for Post Quantum Digital Signatures'', and 2022K48YYP ``Unirationality, Hilbert schemes, and singularities'', and by the MUR Excellence Department Project awarded to Dipartimento di Matematica, Università di Genova, CUP D33C23001110001.

\section{Preliminaries and notation}\label{section:preliminaries}

Let $\Fq$ denote the finite field with $q$ elements and let $\Fq[z_1,\ldots,z_n]$ be the polynomial ring in $n$ variables over $\Fq$.
Given a system $F=\{f_1,\ldots,f_m\}$ of $m$ equations in $\Fq[z_1,\ldots,z_n]$, we denote by $F^{\mathrm{TOP}}$ the system $\{f_1^{\mathrm{TOP}},\ldots,f_m^{\mathrm{TOP}}\}$, where $f_i^{\mathrm{TOP}}$ is the homogeneous component of maximum degree of $f_i$.
Moreover, we denote by $\vs F$ the vector space generated by $F$, by $\ideal F$ the ideal generated by $F$, and by $\ideal F_d$ the vector space of homogeneous elements of degree $d$ of $\ideal F$. 




\subsection{The \pesto{} scheme}

In this section we briefly describe the multivariate  scheme \pesto, proposed by Calderini, Caminata and Villa and provide an overview of the key concepts required. For more detailed information, we refer to the original article~\cite{CCV24}. 

The central map of \pesto{} $F\colon \Fqn \to \Fqm$ is a quadratic polynomial map of the form $F(x,y)=(T(x,y),U(x,y))$, where both $T$ and $U$ have degree at most~$2$.

\begin{definition}[\pesto{}]\label{def:pesto}
    Fix positive integers $n$, $m$, $t$, and $s$ with $t\leq min\{n, m\}$ and $0<s \leq n - t$, then consider the maps:
    \begin{enumerate}
        \item $T\colon \Fq^t\times \Fq^{n-t}\to \Fq^{t}$ defined as $T(x,y)=x+\q(y)$, where $\q\colon \Fq^{n-t}\to \Fq^t$ is random quadratic polynomial map.
        The inverse of $T$ with respect to $x$, denoted as $T_y^{-1}(x)$, is then $x-\q(y)$.
        \item  $U\colon \Fq^t\times \Fq^{n-t}\to \Fq^{m-t}$ a system of $m-t$ random Oil and Vinegar (OV) maps~\cite{Ding2020}, with $x_1,\dots , x_t, y_1, \dots , y_s$ vinegar variables and $y_{s+1}, . . . , y_{n-t}$ oil variables. In other words, in \pesto{} the equations of $U$ have the form
        \begin{equation}\label{eq:U}
            f^{(i)}=\sum_{j,k\in V} \alpha_{jk}^{(i)}z_jz_k+\sum_{j\in V, k\in O}\beta_{jk}^{(i)}z_jz_k+\sum_{j\in V\cup O}\gamma_j^{(i)}z_j+\delta^{(i)}
        \end{equation}
where $V=\{1,\dots,s+t\}$, $O=\{s+t+1,\dots,n\}$, $z_i=x_i$ for $1\leq i\leq t$, $z_i=y_{i-t}$ for $t+1\leq i\leq n$.
The coefficients $\alpha_{jk}^{(i)}, \beta_{jk}^{(i)}, \gamma_j^{(i)}, \delta^{(i)}\in\Fq$ are chosen uniformly at random.
\item $A_1\colon \Fqm \to \Fqm$ a random affine linear bijection.
\item $A_2\colon \Fqn \to \Fqn$ a random affine linear bijection.
    \end{enumerate}
Let $F(x,y)=(x+\q(y),U(x,y))$. The corresponding secret map is the system  $G(x,y)=F(T_y^{-1}(x),U(T_y^{-1}(x),y))=(x - \q(y), U(x - \q(y), y))$. The public key is then given by the map $G_{\mathrm{pub}} = A_1 \circ G \circ A_2$, while the secret key consists of $A_1, A_2, \q, U$.
\end{definition}

We assume throughout the paper that each component of $\q$ has degree 2. In fact, if a component of $\q$ is linear, then composing $x-\q(y)$ with $A_2$ is just a change of coordinates on that component of $G$. However, the goal of applying a CCZ transformation is to introduce a non-linear transformation in addition to~$A_2$. 

Using this setup, one can design either a digital signature scheme or an encryption scheme, as demonstrated by the authors in~\cite[\S4.1.1 and~\S4.1.2]{CCV24}. 
Based on the security analysis performed in~\cite[\S6]{CCV24}, the authors recommended that $t$ should be around $n/3$.

\subsection{Toy example}\label{toy}

We provide a toy example of \pesto{} over $\mathbb{F}_3$, which will be our running example throughout the paper. 
The example was produced using Magma~\cite{BCP97}. The code used to produce it can be found in Appendix~\ref{code:ex}.

We let $n=6$, $m=5$, $t=2$, and $s=1$. Therefore, we have the sets of variables $x=\{x_1, x_2\}$ and $y=\{y_1, y_2, y_3, y_4\}$. In order to define the map $T(x,y)\colon \mathbb{F}_3^{2}\times\mathbb{F}_3^{4} \to \mathbb{F}_3^2 $, we need to choose a random quadratic map $\q\colon \mathbb{F}_3^{4}\to \mathbb{F}_3^2$. Let
\sloppy
\[ 
\q(y) =
\begin{bmatrix}
2y_1^2 + 2y_1 y_2 + 2y_2 y_3 + 2y_3 y_4 + y_4^2 + y_3 + y_4 + 2 \\[0.5cm]
2y_1 y_3 + y_3^2 + y_1 y_4 + 2y_3 y_4 + 2y_4^2 + 2y_1 + 2y_3 + 2
\end{bmatrix}
\]
and let $T(x,y)=x+\q(y)$. The map $U$ is an OV system of $3$ equations with vinegar variables $x_1$, $x_2$, $y_1$ and oil variables $y_2$, $y_3$, $y_4$. We choose
\[ 
U(x,y) =
\begin{bmatrix}
x_1^2 + x_1 x_2 + 2x_1 y_1 + 2x_2 y_1 + 2y_1^2 + x_1 y_2 + x_2 y_2 + y_1 y_2 + 
2x_1 y_3\\ + x_2 y_3 + 2y_1 y_3 + x_1 y_4 + y_1 y_4 + 2x_1 + x_2 + y_1 + y_3 + 1 \\[0.5cm]
x_1 x_2 + 2x_1 y_1 + x_2 y_1 + y_1^2 + 2x_1 y_2 + 2x_2 y_3 + 2x_1 y_4 \\+ x_1 + x_2 + y_1 + 2y_3 + 2 \\[0.5cm]
x_1 x_2 + x_2^2 + 2x_1 y_1 + y_1^2 + x_1 y_2 + y_1 y_2 + 2x_1 y_3 + y_1 y_3 + x_1 y_4\\ + 2x_2 y_4 + y_1 y_4 + 2x_1 + 2y_1 + 2y_2 + 2y_3 + 2y_4 + 1
\end{bmatrix}.
\]
With this choice of $T$ and $U$, $G(x,y)$ takes the form
\[ 
 \begin{bmatrix}
y_1^2 + y_1 y_2 + y_2 y_3 + y_3 y_4 + 2y_4^2 + x_1 + 2y_3 + 2y_4 + 1 \\[0.5cm]
y_1 y_3 + 2y_3^2 + 2y_1 y_4 + y_3 y_4 + y_4^2 + x_2 + y_1 + y_3 + 1 \\[0.5cm]
y_1^4 + 2y_1^3 y_2 + y_1^2 y_2^2 + y_1^3 y_3 + 2y_1 y_2^2 y_3 + 2y_1^2 y_3^2 + y_2^2 y_3^2 + 2y_2 y_3^3 + 2y_1^3 y_4 + 2y_1^2 y_2 y_4 \\
+ 2y_1 y_2 y_3 y_4 + y_1 y_3^2 y_4 + 2y_3^3 y_4 + 2y_1^2 y_4^2 + 2y_1 y_2 y_4^2 + y_1 y_3 y_4^2 + 2y_2 y_3 y_4^2 + y_1 y_4^3 + y_3 y_4^3 \\
+ 2x_1 y_1^2 + x_2 y_1^2 + 2x_1 y_1 y_2 + x_2 y_1 y_2 + y_1^2 y_2 + y_1 y_2^2 + x_1 y_1 y_3 + 2x_1 y_2 y_3 + x_2 y_2 y_3 \\
+ 2y_1 y_2 y_3+ y_2^2 y_3 + 2x_1 y_3^2 + y_1 y_3^2 + 2x_1 y_1 y_4 + y_1 y_2 y_4 + x_2 y_3 y_4 + y_1 y_3 y_4 + y_2 y_3 y_4 \\
+ 2y_3^2 y_4 + 2x_1 y_4^2 + 2x_2 y_4^2 +x_1^2 + x_1 x_2 + 2x_2 y_1 + x_1 y_2 + x_2 y_2 + y_1 y_2 + x_1 y_3 + y_1 y_3 \\
+2y_2 y_3 + 2x_1 y_4 + 2x_2 y_4  +2y_1 y_4 + 2y_2 y_4 + 2y_3 y_4 + 2x_1 + 2x_2 +y_1 + 2y_2 + y_3 + 2y_4\\[0.5cm]
y_1^3 y_3 + y_1^2 y_2 y_3 + 2y_1^2 y_3^2 + 2y_2 y_3^3 + 2y_1^3 y_4 + 2y_1^2 y_2 y_4 + y_1^2 y_3 y_4 + y_1 y_3^2 y_4 + y_2 y_3^2 y_4\\
+ 2y_3^3 y_4 + y_1^2 y_4^2 + y_1 y_2 y_4^2 + y_1 y_3 y_4^2 + y_2 y_3 y_4^2 + 2y_3^2 y_4^2 + y_1 y_4^3 + 2y_4^4+ x_2 y_1^2 + x_2 y_1 y_2\\
+ 2y_1^2 y_2 + 2y_1 y_2^2 + x_1 y_1 y_3 + 2y_1^2 y_3 + x_2 y_2 y_3 + y_1 y_2 y_3+ 2y_2^2 y_3 + 2x_1 y_3^2 + y_2 y_3^2+ 2y_3^3 \\
+ 2x_1 y_1 y_4 + y_1^2 y_4 + 2y_1 y_2 y_4+ x_1 y_3 y_4 + x_2 y_3 y_4 + 2y_1 y_3 y_4 + y_2 y_3 y_4 + x_1 y_4^2 + 2x_2 y_4^2\\
+ 2y_1 y_4^2 + y_2 y_4^2 + y_3 y_4^2+  x_1 x_2 + x_2 y_1 + y_1^2 + 2x_1 y_2 + 2y_1 y_2 + x_1 y_3 + x_2 y_3 + 2y_1 y_3\\
+ 2y_3^2 + 2x_1 y_4 + 2x_2 y_4 + y_1 y_4 + y_2 y_4 + y_3 y_4 + y_4^2+ 2x_1 + 2x_2 + 2y_2 + y_3 + 2\\[0.5cm]
y_1^3 y_3 + y_1^2 y_2 y_3 + y_1 y_3^3 + 2 y_2 y_3^3 + y_3^4 + 
2 y_1^3 y_4 + 2 y_1^2 y_2 y_4 + 2 y_1^2 y_3 y_4 + 2 y_1 y_3^2 y_4\\ + y_2 y_3^2 y_4 + 2 y_1^2 y_4^2 + y_1 y_2 y_4^2 + y_1 y_3 y_4^2 + y_2 y_3 y_4^2 + y_3^2 y_4^2 + 2 y_1 y_4^3+ 2 y_3 y_4^3 + x_2 y_1^2\\
+ x_2 y_1 y_2 + y_1^2 y_2 + y_1 y_2^2 + x_1 y_1 y_3 + 2 x_2 y_1 y_3 + 2 y_1^2 y_3 + x_2 y_2 y_3 + y_2^2 y_3 + 2 x_1 y_3^2 + x_2 y_3^2\\
+ 2 y_1 y_3^2 + 2 y_3^3 + 2 x_1 y_1 y_4 + x_2 y_1 y_4 + 2 y_1^2 y_4 + y_1 y_2 y_4 + x_1 y_3 y_4 + 2 y_1 y_3 y_4 + 2 y_2 y_3 y_4\\
+ x_1 y_4^2 + x_2 y_4^2 + y_1 y_4^2 + 2 y_2 y_4^2 + x_1 x_2 + x_2^2
+ 2 x_2 y_1+ 2 y_1^2 + x_1 y_2 + y_1 y_2 + x_2 y_3 \\
+ 2 y_2 y_3 + y_3^2 + x_1 y_4 + x_2 y_4 + 2 y_2 y_4 + y_3 y_4 + 2 y_4^2 + y_1 + y_3 + 2 y_4 + 2
\end{bmatrix}.
\]
Finally, we choose the following affine linear bijections of  $\mathbb{F}_3^{5}$ and $\mathbb{F}_3^{6}$
\[ 
A_1(x,y) =
\begin{bmatrix}
1 & 0 & 2 & 2 & 0 \\
1 & 2 & 1 & 1 & 1 \\
2 & 2 & 2 & 2 & 0 \\
0 & 1 & 1 & 0 & 1 \\
2 & 2 & 0 & 0 & 0
\end{bmatrix}
\begin{bmatrix}
    z_1 \\ z_2 \\ z_3 \\ z_4 \\ z_5
\end{bmatrix}+
\begin{bmatrix}
    2 \\ 1 \\1 \\ 2 \\ 2
\end{bmatrix},\;\;\;
A_2(x,y) =
\begin{bmatrix}
0 & 2 & 2 & 0 & 2 & 0 \\
0 & 2 & 1 & 2 & 2 & 1 \\
0 & 2 & 2 & 1 & 0 & 0 \\
2 & 1 & 1 & 0 & 0 & 0 \\
1 & 1 & 1 & 2 & 0 & 1 \\
0 & 0 & 1 & 2 & 2 & 2
\end{bmatrix}
\begin{bmatrix}
    x_1 \\ x_2 \\ y_1 \\ y_2 \\ y_3 \\ y_4
\end{bmatrix}+
\begin{bmatrix}
    2 \\ 1 \\2 \\ 1 \\ 0 \\2
\end{bmatrix}.
\]
The functions $\q$, $U$, $A_1$, $A_2$ were randomly generated in Magma~\cite{BCP97}. The public map $G_{\mathrm{pub}} = A_1 \circ G \circ A_2 $ consists of $5$ dense polynomials of degree $4$. Given their sizes, we opt not to report them here. 
To conclude, we choose a random secret input $\mathrm{ipt}=(2,0,1,2,2,0)\in\mathbb{F}_3^6$ for the public system $G_{\mathrm{pub}}$ and we compute the corresponding output $\mathrm{opt}=(1,2,2,1,2)\in\mathbb{F}_3^5$.

\subsection{Security analysis}

In this section, we summarize the security analysis of \pesto{} from~\cite[\S6]{CCV24}. In particular, we focus on some well-known attacks for solving Oil and Vinegar (OV) systems and algebraic attacks using Gröbner Bases, as these attacks will be revisited in the sequel.

\subsubsection{Structural attacks}

The secret map \( F \) is constructed using two Oil and Vinegar systems, \( T(x, y) \) and \( U(x, y) \). This naturally raises the question of whether known attacks on OV systems could also be extended to \pesto{}. After applying the twisting transformation, the first \( t \) coordinates of the map \( G \) retain the structure of an OV system, \( x - \q(y) \), while the remaining \( m - t \) coordinates are usually polynomials of degree $4$. Therefore, if it is possible to isolate the components of \( G_{\mathrm{pub}} \) corresponding to \( x - \q(y) \), the known attacks to OV systems could be directed specifically at these components. In particular, due to the Kipnis–Shamir attack~\cite{KS98} on balanced Oil and Vinegar signature schemes, the authors of~\cite{CCV24} recommend keeping the system \( x - \q(y) \) unbalanced. For instance, setting \( t \approx n/3 \) helps to maintain this imbalance. On the other hand, since the second part of the system \( G \) consists of polynomials with degrees up to 4, they argue that standard OV attacks are unlikely to be effective against the entire map \( G_{\mathrm{pub}} \).
Additionally, they present a linearization attack that can be performed when $s = 0$. Therefore, they recommend choosing $s > 0$.

\subsubsection{Algebraic attacks using Gr\"obner bases}

We consider the scenario where an attacker aims to forge a signature by finding a preimage \( v \in \mathbb{F}_q^n \) of a given element \( w \in \mathbb{F}_q^m \). This can be done by computing a Gröbner basis for the polynomial system \( G_{\mathrm{pub}} \), see e.g.~\cite{CG21} for details on how this can be done. The asymptotic complexity of these methods is 
\[
O\left(\binom{n + \text{sd}(G_{\mathrm{pub}})}{n}^\omega\right),
\]  
where \( n \) is the number of variables, \( \text{sd}(G_{\mathrm{pub}}) \) is the solving degree of the system $G_{\mathrm{pub}}$, and \( 2 < \omega < 3 \) is the matrix-multiplication exponent.  

Informally, the solving degree is the maximum degree of the polynomials produced during the Gr\"obner basis computation, see~\cite{CG23} for a precise definition and more details. In order to estimate the solving degree of the system \( G_{\mathrm{pub}} \) associated with \pesto, the authors of~\cite{CCV24} conducted computational experiments in Magma~\cite{BCP97} for different choices of the parameters \( n, m, t, s, q \). A summary of the results may be found in~\cite[\S6.6]{CCV24}. It appears that the solving degree increases rapidly with respect to the input parameters, making this attack quickly infeasible.

\section{Reducing the public key to a quadratic system}\label{section:STEP1}

In this section, we show how to reduce the public key \( G_{\mathrm{pub}}\) associated to a system of the form $F(x,y)=(T(x,y),U(x,y))$ to a quadratic system, under the assumption that $U,T$, and $T_y^{-1}$ are quadratic. This applies in particular to \pesto, see Definition~\ref{def:pesto}.
We present two methods for carrying on the reduction: one using a linear algebra based algorithm for Gr\"obner bases in~\S\ref{section:method1}, and the other via Higher Order Linearization Equations (HOLE) in~\S\ref{section:method2}.


\subsection{Method 1: Gr\"obner Bases Algorithms}\label{section:method1}

Throughout this subsection, we assume that $T$ has the form $T=x+\q(y)$ as in Definition~\ref{def:pesto}. 
We aim to demonstrate that by running a linear algebra based algorithm on the Macaulay matrix of degree $4$ associated to $G_{\mathrm{pub}}$, we can reduce $G_{\mathrm{pub}}$ to a system of quadratic equations.
To establish this result, we will use the framework of constructible polynomials. We begin by recalling the relevant definitions and some useful properties from \cite{Huang2018207}.

\begin{definition}
    Let $\mathcal{F}\subseteq \mathbb{F}_q[z_1,\dots,z_n]$ be a system of polynomials and let $i\in\mathbb{N}$. 
    With $\mathbb{F}_q[z_1,\dots,z_n]_{\leq i}$ we denote the set of polynomials of degree up to $i$.
    The space $V_{\mathcal{F},i}$ of \emph{constructible polynonomials} is the smallest $\mathbb{F}_q$-vector space such that
    \begin{compactenum}
        \item $\mathcal{F}\cap\mathbb{F}_q[z_1,\dots,z_n]_{\leq i}\subseteq V_{\mathcal{F},i}$;
        \item if $g\in V_{\mathcal{F},i}$ and $f\in \mathbb{F}_q[z_1,\dots,z_n]$ with $\deg(fg)\leq i$ then $fg\in  V_{\mathcal{F},i}$.
    \end{compactenum}
Given two polynomials $f,g$ we write $f\equiv_{\mathcal{F},i}g$ if $f-g\in V_{\mathcal{F},i}$.    
\end{definition}

In the following remark, we collect some properties from \cite[Proposition~2.3]{Huang2018207}, which we will use in the sequel.

\begin{remark}\label{rem_propertiesVf}
Let $\mathcal{F},\mathcal{G}\subseteq \mathbb{F}_q[z_1,\dots,z_n]$ be systems of polynomials, let $i\in\mathbb{N}$, and  let $A:\mathbb{F}_q^n\rightarrow\mathbb{F}_q^n$ be an affine linear bijection. Then, the following properties hold:
\begin{compactenum}
    \item If $\mathcal{F}\subseteq\mathcal{G}$ then $V_{\mathcal{F},i}\subseteq V_{\mathcal{G},i}$;
    \item $V_{\mathcal{F},i}\circ A = V_{\mathcal{F}\circ A,i}$.
\end{compactenum}
\end{remark}

\begin{lemma}\label{lemma.1}\label{lemma.1bis}
Let $U\subseteq\Fq[x_1,\dots,x_t,y_1,\dots,y_{n-t}]$ and $\q=\{\q_1,\dots,\q_t\}\subseteq\Fq[y_1,\dots,y_{n-t}]$ be systems of quadratic equations and let $u$ be an equation of the system $U(x - \q(y), y))$. 
Then $u\equiv_{x-\q(y),4} v$ with $\deg v\leq 2$.  
\end{lemma}

\begin{proof}
Assuming that the $i$-th equation of $U(x,y)$ has the form
\begin{align*}
    \sum_{j=1}^t\sum_{k=j}^t\alpha_{j,k}x_jx_k+\sum_{j=1}^t\sum_{k=1}^{n-t}\beta_{j,k}x_jy_k+\sum_{j=1}^{n-t}\sum_{k=j}^{n-t}\gamma_{j,k}y_jy_k+\sum_{j=1}^t\delta_{j}x_j+\sum_{j=1}^{n-t}\epsilon_{j}y_j+\zeta,
\end{align*}
the $i$-th equation of $U(x-\q(y),y)$ is of the form
\begin{align*}
 u= & \sum_{j=1}^t\sum_{k=j}^t\alpha_{j,k}x_jx_k-
    \sum_{j=1}^t\sum_{k=j}^t\alpha_{j,k}x_j\q_k-
    \sum_{j=1}^t\sum_{k=j}^t\alpha_{j,k}\q_jx_k+
    \sum_{j=1}^t\sum_{k=j}^t\alpha_{j,k}\q_j\q_k\\
    &+
   \sum_{j=1}^t\sum_{k=1}^{n-t}\beta_{j,k}x_jy_k-\sum_{j=1}^t\sum_{k=1}^{n-t}\beta_{j,k}\q_jy_k+\sum_{j=1}^{n-t}\sum_{k=j}^{n-t}\gamma_{j,k}y_jy_k+\sum_{j=1}^t\delta_{j}x_j\\
   &-\sum_{j=1}^t\delta_{j}\q_j+\sum_{j=1}^{n-t}\epsilon_{j}y_j+\zeta,
\end{align*}
where we removed the dependency on $i$ from the coefficients and on the input variables $y$ from the $\q_j$'s to ease the notation.
We denote by $u^r$ and $\q_j^r$ the homogeneous component of degree $r$ of $u$ and $\q_j$, respectively.
Therefore, we have
\begin{align*}
 u^4=  &  \sum_{j=1}^t\sum_{k=j}^t\alpha_{j,k}\q_j^2\q_k^2\\
 u^3=&\sum_{j=1}^t\sum_{k=j}^t\alpha_{j,k}\q_j^2\q_k^1+\sum_{j=1}^t\sum_{k=j}^t\alpha_{j,k}\q_j^1\q_k^2
 -
    \sum_{j=1}^t\sum_{k=j}^t\alpha_{j,k}x_j\q^2_k\\
    &-
    \sum_{j=1}^t\sum_{k=j}^t\alpha_{j,k}\q_j^2x_k
   -\sum_{j=1}^t\sum_{k=1}^{n-t}\beta_{j,k}\q^2_jy_k
   =\sum_{j=1}^t\q_j^2h_{j}^1(x,y),
\end{align*}
where $h_{j}^1(x,y)$ is a suitable homogeneous linear polynomial, for each $j=1,\ldots,t$.
Let $p_j=\sum_{k=j}^t\alpha_{j,k}\q_k^2+h_{j}^1(x,y).$
Then $u^4+u^3=\sum_{j=1}^tp_j\q_j^2$ and
\begin{align*}
p_j = & \sum_{k=j}^t\alpha_{j,k}\left(\q_k^2+\q_k^1-x_k\right)+\sum_{k=1}^j\alpha_{k,j}\left(\q_k^1-x_k\right)-\sum_{k=1}^{n-t}\beta_{j,k}y_k \\
= & \sum_{k=j}^t\alpha_{j,k}\left(\q_k(y)-x_k-\q_k^0\right)+\sum_{k=1}^j\alpha_{k,j}\left(\q_k^1-x_k\right)-\sum_{k=1}^{n-t}\beta_{j,k}y_k 
\equiv_{x-\q(y),2}\ell_j,
\end{align*} where $\deg(\ell_j)\leq 1$.
Hence 
\begin{align*}
u\equiv_{x-\q(y),4} & u+\sum_{j=1}^t p_j(x_j-\q_j)=u^2+u^1+u^0+\sum_{j=1}^t p_j(x_j-\q_j^1-\q_j^0)\\ \equiv_{x-\q(y),4} & u^2+u^1+u^0+\sum_{j=1}^t \ell_j(x_j-\q_j^1-\q_j^0),
\end{align*} 
an element of degree up to $2$.
\end{proof}

We are now ready to present the reduction of the public system $G_{\mathrm{pub}}$ of \pesto{} to a quadratic system, which we formalize in the next theorem.

\begin{theorem}\label{thm.1}
Consider $T:\Fqn\to\fq^t$ such that $T(x,y)=x+\q(y)$, where $\q(y)$ is quadratic. Consider $U:\Fqn\to\fq^{m-t}$ a system of quadratic equations. For $G=(T_y^{-1}(x),U(T_y^{-1}(x),y))$, set $G_{pub}=A_1\circ G\circ A_2$, with $A_1,A_2$ linear affine bijections of $\Fqm$ and $\Fqn$ respectively.
Starting from the equations of $G_{\mathrm{pub}}$ and $c\in\Fqm$, we can produce:
\begin{enumerate}[1)]
    \item a linear space of quadratic equations which contains the equations of $(x-\q(y))\circ A_2$ with complexity $\mathcal{O}(m^2n^4)$,
    \item a quadratic system with the same solutions as $G_{\mathrm{pub}}=c$ with complexity $\mathcal{O}(n^{16})$,
\end{enumerate}
\end{theorem}

\begin{proof}
\textit{1)} Let $W$ be the subspace of $\vs{G_{\mathrm{pub}}}$ consisting of the equations of degree at most $2$. Since $\vs{G_{\mathrm{pub}}}=\langle G\circ A_2\rangle$, $W$ contains the $t$ linearly independent quadratic equations of $(x-\q(y))\circ A_2$.

Build a matrix $M$ associated to the system $G_{\mathrm{pub}}$ with respect to a degree-compatible term order as follows. Its columns correspond to the monomials of degree up to $4$ in decreasing order with respect to the chosen order. In particular, each column corresponds to a monomial of degree smaller than or equal degree to the monomial corresponding to the column on its left. Each row corresponds to one of the equations of $G_{\mathrm{pub}}$ and the entry in the row corresponding to $g$ and in the column corresponding to the monomial $\eta$ is the coefficient of $\eta$ in $g$. Therefore, there is a natural identification between the rowspace of $M$ and $\vs{G_{\mathrm{pub}}}$. 

Put $M$ in reduced row-echelon form. Notice that now $M$ has the property that each row corresponds to a polynomial of degree smaller than or equal to the polynomial corresponding to the row above. Since $M$ is in reduced row-echelon form, $W$ is the vector space generated by the rows which correspond to polynomials of degree at most $2$, showing that the last $\dim(W)$ rows of $M$ are a basis of $W$.
The complexity of computing it is $\mathcal{O}(m^2n^4)$, i.e., the complexity of Gaussian elimination in $M$.

\textit{2)} For ease of notation and up to incorporating the constant in the equations of $G_{\mathrm{pub}}$, we may assume without loss of generality that $c=0$. To produce a system of quadratic equations with the same solutions as $G_{\mathrm{pub}}$,  
we consider the Macaulay matrix $\mathcal{M}$ of degree $4$ associated to the system $G_{\mathrm{pub}}$ with respect to a degree-compatible term order. Notice that the matrix $M$ constructed in \textit{1)} is a submatrix of $\mathcal{M}$. 
Recall that $\mathcal{M}$ is built as follows. Its columns correspond to the monomials of degree up to $4$ ordered in decreasing order with respect to the chosen order and the rows correspond to all the equations $\mu g$, where $g$ is one the equations of $G_{\mathrm{pub}}$ and $\mu$ is a monomial such that $\deg(\mu g)\leq 4$. The entry in the row corresponding to $\mu g$ and in the column corresponding to the monomial $\eta$ is the coefficient of $\eta$ in $\mu g$. Throughout the proof, we abuse terminology and we identify the rows of the matrix with the corresponding polynomials. 

We run a Gr\"obner basis-type computation on $\mathcal{M}$ by computing its reduced row-echelon form, together with the mutant trick. More precisely, we perform Gaussian elimination on $\mathcal{M}$ and, whenever a new polynomial $f$ with $\deg(f)<4$ appears, then we append to the matrix new rows corresponding to $\mu f$, for every monomial $\mu$ such that $\deg(\mu f)\leq 4$. Then we continue with Gaussian elimination on $\mathcal{M}$. This procedure eventually terminates, since the number of rows in the reduced row-echelon form of the resulting matrix is bounded from above by the number of monomials of degree up to $4$. It follows from the proof of~\cite[Proposition~3.13]{Sal23} that the complexity of this computation is $\mathcal{O}(n^{16})$, since $\mathcal{M}$ has $\mathcal{O}(n^4)$ columns.

We claim that the procedure described above produces a system, which generates the same ideal as $G_{\mathrm{pub}}$. This follows from the fact that computing the reduced row-echelon form of a matrix does not change the vector space, hence the ideal, generated by its rows. Moreover, appending to a matrix rows which are multiples of other rows does not change the ideal generated by its rows.
In particular, the algorithm described produces a system which has the same solutions as $G_{\mathrm{pub}}$. 
Since the term order is degree-compatible, by \cite[Theorem~1]{GMP22} the rowspace of the Macaulay matrix coincides with the vector space $V_{G_{\mathrm{pub}},4}$ of constructible polynomials of degree $4$ of $G_{\mathrm{pub}}$. 

We conclude by proving that the system produced is quadratic.
Consider any row $g$ of $\mathcal{M}$ which has degree $3$ or $4$ after the first round of Gaussian elimination. We claim that $g\equiv_{G_{\mathrm{pub}},4} g'$, for some $g'$ with $\deg(g')\leq2$.
In \textit{1)} we showed that, after the first round of Gaussian elimination, the matrix contains rows which generate all the elements of $\langle(x-\q(y))\circ A_2\rangle$.
Thus, by Remark~\ref{rem_propertiesVf} we have $V_{(x-\q(y))\circ A_2,4}\subseteq V_{G_{\mathrm{pub}},4}$,
so it suffices to show that $g\equiv_{(x-\q(y))\circ A_2,4} g'$. Equivalently, it suffices to show that $g\circ A_2^{-1}\equiv_{(x-\q(y)),4} g'\circ A_2^{-1}$, since an affine linear bijection preserves the degree. Since $g\circ A_2^{-1}$ is a linear combination of the equations of $U$, we conclude by Lemma~\ref{lemma.1}.
\end{proof}

\begin{example}
Running the algorithm from Theorem~\ref{thm.1},~\textit{2)} on the Toy Example from~\S\ref{toy} yields the following reduced quadratic public system.
\end{example}
\[ 
G_{\mathrm{red}}= \begin{bmatrix}
2 x_1 x_2 + 2 x_1 y_1 + x_2 y_1 + 2 x_2 y_2 + 2 x_2 y_3 + y_1 y_3 + 2 y_2 y_3 + 2 y_3^2 + 2 x_1 y_4 \\
+ x_2 y_4 + 2 y_2 y_4 + y_3 y_4 + y_4^2 + 2 x_1 + x_2 + y_1 + y_2 + 2 y_3 \\[0.5cm]
2 x_1^2 + 2 x_1 x_2 + x_2^2 + 2 x_2 y_1 + y_1^2 + 2 x_1 y_2 + x_2 y_2 + y_1 y_2 + y_2^2 + 2 x_1 y_3\\
+ 2 x_1 y_4 + 2 x_2 y_4 + 2 y_2 y_4 + 2 y_3 y_4 + x_1 + y_1 + y_2 + 2 y_3 + y_4 + 2 \\[0.5cm]
2 x_2^2 + x_2 y_1 + x_1 y_2 + y_1 y_2 + 2 y_2^2 + y_1 y_3 + 
2 y_2 y_3 + y_3^2 + x_1 y_4 + x_2 y_4 \\
+ 2 y_1 y_4 + 2 y_2 y_4 + 2 y_3 y_4 + 2 y_4^2 + x_1 + 2 y_1 + 2 y_2 + 2 y_3 + y_4 + 2\\[0.5cm]
2 x_2 y_1 + y_1^2 + 2 x_1 y_2 + 2 x_2 y_2 + 2 y_1 y_2 + x_2 y_3 + 2 y_3^2 + 2 x_1 y_4 \\
+ 2 x_2 y_4 + y_1 y_4 + y_2 y_4 + y_3 y_4 + y_4^2 + y_1 + y_3 + 2 y_4 \\[0.5cm]
x_2^2 + 2 x_2 y_1 + 2 y_1^2 + x_1 y_2 + 2 x_2 y_2 + y_2^2 + y_1 y_3 + 2 y_2 y_3 + y_3^2 + x_1 y_4\\
+ x_2 y_4 + 2 y_1 y_4 + 2 y_2 y_4 + 2 y_3 y_4 + 2 y_4^2 + x_2 + y_1 + y_2 + y_3 + 1
\end{bmatrix}.
\]

\subsection{Method 2: A Higher Order Linearization Equation (HOLE) attack}\label{section:method2}

In this section, we recall the linearization attack and we derive a similar method to reduce the public system to a system of quadratic equations. 

\subsubsection{Higher order linearization equations (HOLEs)}

\begin{definition}
    Consider $\mathcal P=(p_1,\ldots,p_m):\Fqn\to\Fqm$ the public key of a multivariate public key primitive, with $p_i\in\fq[x_1,\ldots,x_n]$ for $1\le i\le m$. \\
    A \emph{(valid) input/output pair} is a pair $(\Bar{z},\Bar{w})\in\fq^n\times\fq^m$ such that $\mathcal P(\Bar{z})=\Bar{w}$. \\
    A \emph{linearization equation} is a polynomial equation in 
    $\fq[z_1,\ldots,z_n,w_1,\ldots,w_m]$ of the form
    \begin{equation}\label{eq:linearization eq}
        \mathcal R(z,w)=\sum_{i=1}^m\sum_{j=1}^n\alpha_{ij}w_iz_j+\sum_{i=1}^m\beta_iw_i+\sum_{j=1}^n\gamma_jz_j+\delta,
    \end{equation}
    such that $\mathcal R(\Bar z,\Bar w)=0$ for any valid input/output pair $(\Bar z,\Bar w)$.\\
     A \emph{higher order linearization equation (HOLE)} is a polynomial equation  of the form
    \begin{equation}\label{eq:HOLE}
        \mathcal R(z,w)=\sum_{j=1}^ng_j(w_1,\ldots,w_m)z_j+g(w_1,\ldots,w_m),
    \end{equation}
    with $g,g_j\in\fq[w_1,\ldots,w_m]$ for $1\le j\le n$, such that $\mathcal R(\Bar z,\Bar w)=0$ for any valid input/output pair $(\Bar z,\Bar w)$.
    The \emph{degree} of the HOLE $\mathcal R$ is given by
    $$d=\max\{\deg(g_1),\ldots,\deg(g_n),\deg(g)\}.$$
\end{definition}

Notice that, when evaluating \eqref{eq:linearization eq} and \eqref{eq:HOLE} at a ciphertext (output) $(\Bar w_1,\ldots,\Bar w_m)$, one obtains linear equations in the plaintext (input) variables $z_1,\ldots,z_n$, that is, equations of the form $\mathcal R(z,\Bar w)=0$.
\\

The existence of  linearization equations for a public key $\mathcal P$ allows to perform the so-called \emph{linearization equation attack}. This attack was used  by Patarin in~\cite{Pat95}  to break the Matsumoto-Imai cryptosystem~\cite{MI}, one of the first multivariate public-key cryptosystem proposed.

The idea of the attack is the following.
Suppose that we have a public key~$\mathcal P$ and one knows that it satisfies some linearization equations as in \eqref{eq:linearization eq}. Then by computing enough input/output pairs and substituting them into the linearization equation, one gets a system of linear equations in the coefficients $\alpha_{ij},\beta_i,\gamma_j$ and $\delta$, which can be solved by Gaussian elimination.
In case of HOLEs as in \eqref{eq:HOLE}, the number of needed input/output pairs increases exponentially with the degree $d$.
For example, for $d=1$ one needs approximately $(m+1)\cdot(n+1)$, while for $d=2$ one needs approximately $(n+1)\cdot\frac{(m+1)(m+2)}{2}$ input/output pairs.

After performing Gaussian elimination, one gets a system of equations in the input/output variables $z_1,\ldots,z_n,w_1,\ldots,w_m$.
Now, assume that one wishes to decrypt a ciphertext $\Bar{w}=(\Bar w_1,\ldots,\Bar w_m)$.
By substituting these values into the previously computed equations, one obtains linear equations in the plaintext variables $z_1,\ldots,z_n$.
Sometimes, one gets enough linear equations to recover the plaintext. Else, the linear equations can be used to compute partial information on the input, which in turn may be used to speed up other attacks against the cryptosystem.

\subsubsection{Recovering more quadratic equations with the HOLE-technique}

We briefly recall the setup from Section \ref{section:preliminaries}.
For $T:\Fqn\to\fq^t$ such that $T(x,y)=T_y(x)$ is invertible for any fixed $y$, and $U:\Fqn\to\fq^{m-t}$ quadratic, we let
\begin{align*}
    F(x,y)=&\left( T_y(x), U(x,y) \right),\\
    G(x,y)=&\left( T_y^{-1}(x), U(T_y^{-1}(x),y) \right).
\end{align*}
Notice that for $G(x,y)=(w_T,w_U)\in\fq^t\times\fq^{m-t}$, it holds $w_U=U(w_T,y)$.

Consider an arbitrary quadratic system $U(x,y)$ of $m-t$ polynomials of the form
$$\sum_{j=1}^t\sum_{k=j}^t\alpha^{(i)}_{j,k}x_jx_k+\sum_{j=1}^t\sum_{k=1}^{n-t}\beta^{(i)}_{j,k}x_jy_k+\sum_{j=1}^{n-t}\sum_{k=j}^{n-t}\gamma^{(i)}_{j,k}y_jy_k+\sum_{j=1}^t\delta^{(i)}_{j}x_j+\sum_{j=1}^{n-t}\epsilon^{(i)}_{j}y_j+\zeta^{(i)},$$
for $1\le i\le m-t$.
We introduce the input and output variables $x=(z_1,\ldots,z_t)$, $y=(z_{t+1},\ldots,z_n)$, $w_T=(w_1,\ldots,w_t)$ and $w_U=(w_{t+1},\ldots,w_m)$. The relation $w_U=U(w_T,y)$ translates into the quadratic relations
\begin{align*}
w_{t+i}= & \sum_{j=1}^t\sum_{k=j}^t\alpha^{(i)}_{j,k}w_jw_k+\sum_{j=1}^t\sum_{k=1}^{n-t}\beta^{(i)}_{j,k}w_jz_{k+t}+\sum_{j=1}^{n-t}\sum_{k=j}^{n-t}\gamma^{(i)}_{j,k}z_{j+t}z_{k+t}\\
 & +\sum_{j=1}^t\delta^{(i)}_{j}w_j+\sum_{j=1}^{n-t}\epsilon^{(i)}_{j}z_{j+t}+\zeta^{(i)},
 \end{align*}
for $1\le i\le m-t$. Let
\begin{align}\label{eqn4}
    \mathcal R_G^{(i)}(z,w)=&\sum_{j=1}^t\sum_{k=j}^t\alpha^{(i)}_{j,k}w_jw_k+\sum_{j=1}^t\sum_{k=1}^{n-t}\beta^{(i)}_{j,k}w_jz_{k+t}+\sum_{j=1}^{n-t}\sum_{k=j}^{n-t}\gamma^{(i)}_{j,k}z_{j+t}z_{k+t}\nonumber\\
    &+\sum_{j=1}^t\delta^{(i)}_{j}w_j+\sum_{j=1}^{n-t}\epsilon^{(i)}_{j}z_{j+t}+\zeta^{(i)}-w_{i+t},
\end{align}
for $1\le i\le m-t$. The equations (\ref{eqn4}) satisfy $\mathcal R_G^{(i)}(\Bar z,\Bar w)=0$ for $1\le i\le m-t$ and any input/output pair $(\Bar z,\Bar w)$ such that $G(\Bar z)=\Bar w$.

\begin{remark}
     Notice that the $m-t$ relations $\mathcal R_G^{(1)}(z,w),\ldots,\mathcal R_G^{(m-t)}(z,w)$ are linearly independent, since $\mathcal R_G^{(i)}(z,w)$ is the only one among them that involves the variable $w_{i+t}$.
\end{remark}

Since $G_{\mathrm{pub}}$ is obtained from $G$ by composing with affine linear bijections, we have  $m-t$ linearly independent quadratic relations also for the public key. These relations take the form
\begin{align}\label{eq:general relation}
    \mathcal R_{pub}^{(i)}(z,w)=&\sum_{j=1}^m\sum_{k=j}^m\Bar\alpha_{j,k}^{(i)}w_jw_k+\sum_{j=1}^m\sum_{k=1}^{n}\Bar\beta^{(i)}_{j,k}w_jz_{k}+\sum_{j=1}^n\sum_{k=j}^{n}\Bar\gamma^{(i)}_{j,k}z_{j}z_{k}\nonumber\\
    &+\sum_{j=1}^m\Bar\delta^{(i)}_{j}w_j+\sum_{j=1}^{n}\Bar\epsilon^{(i)}_{j}z_{j}+\Bar\zeta,
\end{align}
for $i=1,\ldots,m-t$.

In order to recover the coefficients of $\mathcal R_{pub}^{(i)}(w,z)$, we need to compute around ${n+m+2\choose 2}$ input/output pairs and then perform Gaussian elimination on a matrix of approximately the same size.

Once the  relations $\mathcal R_{pub}^{(i)}(z,w)$ are found, we evaluate all relations at the target value $\Bar w$, obtaining the relations in the input variables
$$\mathcal R_{pub}^{(1)}(z,\Bar w),\dots,\mathcal R_{pub}^{(m-t)}(z,\Bar w).$$

\begin{remark}
We stress that the effectiveness of using HOLE-techniques to recover more quadratic equations does not depend on a specific choice of $U$ and $T$.
We only assume that $U$ is a system of quadratic equations.
No assumption on $T$ is needed, beyond being invertible. In the sequel, we analyze in more detail the case when $T_y^{-1}(x)$ has degree two.
\end{remark}


If $T$ is such that $T_y^{-1}(x)$ is quadratic, then the relation $w_T=T_y^{-1}(x)$ yields $t$ linearly independent quadratic relations between the input and the output of $G$, implying the existence of other $t$ linearly independent quadratic relations between the input and the output of $G_{\mathrm{pub}}$. The equations are linearly independent also from the ones that we previously found, since they do not involve $w_{t+1},\dots,w_m$ and the $i$-th among these additional equations is the only one involving $w_i$ for $1\leq i\leq m$.
Therefore, when performing the Gaussian elimination on the matrix used to recover the relations \eqref{eq:general relation}, we obtain a set of solutions of dimension at least $m$.
Notice that this is the case for \pesto, where $T(x,y)=x+\q(y)$ and $T_y^{-1}(x)=x-\q(y)$, where $\q:\fq^{n-t}\to\fq^t$ is quadratic.

We now explain how to isolate within this set of relations those that correspond to the first $t$ equations of $G$, i.e., to $w_T=T_y^{-1}(x)$. Since $T_y^{-1}(x)$ is quadratic, $w_T=T_y^{-1}(x)$ corresponds to $t$ relations of the form 
\begin{equation}\label{eqn:shortereq}
\sum_{j,k=1}^n\Bar\gamma_{j,k}z_{j}z_{k}+\sum_{j=1}^m\Bar\delta_{j}w_j+\sum_{j=1}^{n}\Bar\epsilon_{j}z_{j}+\Bar\zeta,    
\end{equation} 
where $z_1,\dots,z_n$ are the input variables and $w_1,\dots,w_m$ the output variables.
Since $G$ and $G_{pub}$ are related by two linear affine bijections, one in the input and one in the output variables, the input/output pairs of $G_{pub}$ also satisfy $t$ relations of the form~(\ref{eqn:shortereq}).
Therefore, the linear space of relations~(\ref{eq:general relation}) for $G_{pub}$ contains at least $t$ relations of the form~(\ref{eqn:shortereq}):
$$\mathcal R_{pub}^{(i)}(z,w)=\sum_{j=1}^n\sum_{k=j}^{n}\Bar\gamma^{(i)}_{j,k}z_{j}z_{k}+\sum_{j=1}^m\Bar\delta^{(i)}_{j}w_j+\sum_{j=1}^{n}\Bar\epsilon^{(i)}_{j}z_{j}+\Bar\zeta.$$
In order to find them, one can take a basis of the space of the quadratic relations of $G_{pub}$ and write its elements as rows of a matrix, whose columns correspond to the monomials of degree up to $2$ in $z,w$ ordered in degree-lexicographic or degree-reverse-lexicographic order with $w_1>\dots>w_m>z_1>\dots>z_n$. Since the row space of the matrix contains a vector space of dimension $d\geq t$ of equations which do not involve any of the monomials $w_jw_k$ and $w_jz_k$, the last $d$ rows of the reduced row-echelon form of the matrix generate the vector subspace of relations of $G_{pub}$ of the form~(\ref{eq:general relation}), which contains the desired $t$ relations. Notice that, as in the Gr\"obner basis approach, if $d>t$, one may proceed with the $d$ equations that were found and is still guaranteed to be able to reduce the system $G_{pub}$ to a quadratic system.

\begin{theorem}\label{thm.2}
Consider $T:\Fqn\to\fq^t$ such that $T(x,y)=T_y(x)$ is invertible for any fixed $y$ and $T_y^{-1}(x)$ is quadratic. Consider $U:\Fqn\to\fq^{m-t}$ a system of quadratic equations. For $G=(T_y^{-1}(x),U(T_y^{-1}(x),y))$, set $G_{pub}=A_1\circ G\circ A_2$, with $A_1,A_2$ linear affine bijections of $\Fqm$ and $\Fqn$ respectively.
Starting from the equations of $G_{pub}$ and $c\in\mathbb{F}_q^m$, we can produce:
\begin{enumerate}[1)]
        \item a system of quadratic relations for $G_{pub}$ with complexity $\mathcal O(\max\{mn^6,m^3n^4,m^6\})$,
        \item a quadratic system with the same solutions as $G_{pub}=c$ with additional negligible complexity.
\end{enumerate}
\end{theorem}
\begin{proof}
\textit{1)} The system of quadratic relations for $G_{pub}$, introducing variables also for the output space, can be obtained via Method 2, as already explained.
We recall that the method consists of the following steps:\\
$a)$ compute around ${n+m+2\choose 2}$ random evaluation of the public system;\\
$b)$ perform Gaussian elimination on the corresponding matrix to recover the coefficients of the quadratic relations.\\
The complexity of evaluating the public system at a single chosen input is $\mathcal{O}(m{n+4\choose 4})$, since we have $m$ equations of degree 4 in $n$ variables.
This has to be done about ${n+m+2\choose 2}$ times, in order to produce the system to be solved.
The complexity of solving the linear system of approximate size ${n+m+2 \choose 2}$ is $\mathcal{O}({n+m+2 \choose 2}^3)$. 
Therefore, the complexity for this method is
$\mathcal{O}(m{n+4\choose 4}{n+m+2\choose 2} +{n+m+2 \choose 2}^3)=\mathcal O(mn^6+m^3n^4+m^6)$.\\ 
\textit{2)} To produce the desired quadratic system, we evaluate the output variables at $c$. Hence we evaluate ${m+2 \choose 2}+mn$ terms. The complexity of this final operation is $\mathcal O(m{m+2 \choose 2}+m^2\cdot n)=\mathcal{O}(m^3+m^2n)$.
\end{proof}

\begin{example}
Performing the attack explained in Theorem~\ref{thm.2} on the Toy Example from~\S\ref{toy} yields the reduced quadratic public system
\[ 
G_{\mathrm{red}}= \begin{bmatrix}
y_1^2 + x_1 y_2 + x_2 y_2 + 2 y_1 y_2 + y_1 y_3 + 2 y_2 y_3 + y_3^2+ x_1 y_4 + x_2 y_4 \\
+ 2 y_1 y_4+ 2 y_2 y_4 + 2 y_3 y_4 + 2 y_4^2 +2 x_1 + 2 x_2 + 2 y_4 \\[0.5cm]
2 x_2 y_1 + x_1 y_2 + x_2 y_2 + x_2 y_3 + 2 y_1 y_3 + y_2 y_3 + 
y_3^2 + x_1 y_4 + x_2 y_4\\
+ 2 y_1 y_4 + 2 y_2 y_4 + 2 y_3 y_4 + 2 y_4^2 + x_1 + x_2 + y_1 + y_3 \\[0.5cm]
x_2^2 + 2 x_2 y_1 + 2 y_1^2 + x_1 y_2 + 2 x_2 y_2 + y_2^2 + y_1 y_3 + 2 y_2 y_3 + y_3^2 + x_1 y_4 \\
+ x_2 y_4 + 2 y_1 y_4 + 2 y_2 y_4 + 2 y_3 y_4 + 2 y_4^2 + x_2 + y_1 + y_2 + y_3 + y_4 + 1 \\[0.5cm]
x_1^2 + 2 x_2^2 + 2 x_1 y_1 + 2 x_2 y_1 + 2 y_1^2 + x_1 y_2 + x_2 y_2+ 2 y_1 y_2 + 2 y_2^2+ x_1 y_3\\ 
+ 2 x_2 y_3 + y_1 y_3 + 2 y_2 y_3 + 2 y_3^2 + 2 x_2 y_4 + 2 y_3 y_4 + y_4^2 + x_1 + x_2 + 2 y_4 + 1 \\[0.5cm]
x_1 x_2 + x_1 y_1 + 2 x_2 y_1 + x_2 y_2 + x_2 y_3 + 2 y_1 y_3 + 
y_2 y_3 + y_3^2 + x_1 y_4 \\
+ 2 x_2 y_4 + y_2 y_4 + 2 y_3 y_4 + 2 y_4^2 + x_1 + 2 x_2 + 2 y_1 + 2 y_2 + y_3
\end{bmatrix}.
\]
\end{example}

In the following remark, we point out that for the $\pesto$ scheme the quadratic system obtained as output of Theorem~\ref{thm.2} \textit{2)} is affine equivalent to an (unbalanced) Oil and Vinegar system.
\begin{remark}
We keep the same notations as in Theorem~\ref{thm.2} and we denote by $z=(x,y)$ the input variables, by $w=(w_T,w_U)$ the output variables, and by $S(x,y)=T_y^{-1}(x)$.
By applying the HOLE technique, the relations we get for $G$ are a linear combination of $w_T-S(x,y)=0$ and $w_U-U(w_T,y)=0$.
Now, since $G_{\mathrm{pub}}=A_1\circ G \circ A_2$, setting $w=G_{\mathrm{pub}}(z)$ we obtain $A_1^{-1}(w)=G(A_2(z))$.
This corresponds to applying the affine linear bijection $A_2$ to the input variables and the affine linear bijection $A_1^{-1}$ to the output variables. We split each of these linear maps into two components $A_{1,S}^{-1}:\mathbb{F}_q^{t}\times\mathbb{F}_q^{m-t}\rightarrow\mathbb{F}_q^t$ $A_{1,U}^{-1}:\mathbb{F}_q^{t}\times\mathbb{F}_q^{m-t}\rightarrow\mathbb{F}_q^{m-t}$,$A_{2,S}:\mathbb{F}_q^{t}\times\mathbb{F}_q^{n-t}\rightarrow\mathbb{F}_q^t$ $A_{2,U}:\mathbb{F}_q^{t}\times\mathbb{F}_q^{n-t}\rightarrow\mathbb{F}_q^{n-t}$
Thus, for $G_{\mathrm{pub}}$ the relations that we obtain via HOLE technique are a linear combination of 
\[
\begin{cases}
A_{1,S}^{-1}(w)-S\circ A_2(z)=0\\
A_{1,U}^{-1}(w)-U(A_{1,S}^{-1}(w),A_{2,U}(z))=0
\end{cases}
\]
The last step of Method~2 is to evaluate the target output $\Bar{w}\in\mathbb{F}_q^m$ in the previous system and solving with respect to $z$.
This amounts to solving a quadratic system of the form
\[
\begin{cases}
S\circ A_2(z)=A_{1,S}^{-1}(\Bar{w})\\
U(A_{1,S}^{-1}(\Bar{w}),A_{2,U}(z))=A_{1,U}^{-1}(\Bar{w})
\end{cases}
\]
So, if $S$ and $U$ are (unbalanced) Oil and Vinegar polynomial maps (as in the case of $\pesto$), then the previous system is affine equivalent to an (unbalanced) OV system.
\end{remark}

\section{Conclusions}\label{section:STEP2}

Using either Method 1 or Method 2, we can reduce the public system of $\pesto$, originally consisting of quartics, to a quadratic system. 
More in general, we can reduce to a quadratic system the quartic public system obtained by disguising a quadratic central function using a CCZ transform, as proposed in~\cite{CCV24}. 


Once we obtain a quadratic system via our methods, however, we still must solve it to retrieve the input. 
If the obtained quadratic system is a disguised OV system, as it is the case for the system obtained by applying Method 2 to \pesto, then specific attacks for OV can be used. Otherwise generic attacks, such as those based on Gr\"obner basis techniques, may be used. One may also try to use taylored attacks which exploit the structure of the central function. We do not however know of a general way of translating an attack on a multivariate public scheme to a scheme where the same central function is disguised using a CCZ transform. 

We conclude that the system \pesto{} proposed in~\cite{CCV24} is no more secure than the OV system on which it is based. For multivariate schemes with a quadratic central function disguised by means of a CCZ transform, we believe that the increase in degree when producing a public system from a private system via the CCZ transform does not offer an additional layer of security. In fact, we are able to reduce the quartic public system back to a quadratic system. The question of whether disguising a central function via the CCZ transform may make uneffective certain attacks that work when disguising the central function using linear affine transformations remains open.

\begin{appendix}
    
    \section{Implementation Details}
    
    We provide the Magma~\cite{BCP97} code used to implement the toy example and to reduce to a quadratic system with both Method 1 and Method 2.
    
    \subsection{The Code to Generate the Public System of \pesto}\label{code:ex}
    
    \begin{verbatim}
// chosen parameters
n:=6;  m:=5; t:=2; s:=1;  q:=3; 

Fq:=GF(q); // finite field
Vn:=VectorSpace(Fq,n); // input space
Vm:=VectorSpace(Fq,m); // output space
P<[x]>:=PolynomialRing(Fq,n,"grevlex");
Q<[z]>:=PolynomialRing(Fq,m,"grevlex");
GLn:=GL(n,Fq);  GLm:=GL(m,Fq);

// construct q(y) (t quadratic equations in n-t variables)
qq:=[];	 	
for ind in [1..t] do
    f:=&+[x[i1]*x[i2]*Random(Fq) : i1,i2 in [t+1..n] | i1 le i2]+
        &+[x[i1]*Random(Fq) : i1 in [t+1..n]]+Random(Fq);
    qq:=Append(qq,f);
end for;

// construct U(x,y) (m-t quadratic equations with t+s variables vinegar and n-t-s oil)
U:=[];	 
for ind in [1..m-t] do
    u:=&+[x[i1]*x[i2]*Random(Fq) : i1,i2 in [1..t+s] | i1 le i2]+
        &+[x[i1]*x[i2]*Random(Fq) : i1 in [1..t+s], i2 in [t+s+1..n]]+
        &+[x[i1]*Random(Fq) : i1 in [1..n]]+Random(Fq);
    U:=Append(U,u); 
end for;

// compute U(x-q(y),y)
Ug:=[];	 
for j in [1..m-t] do
    ug:= Evaluate(U[j],[x[i]-qq[i] : i in [1..t]] cat [x[i] : i in [t+1..n]]);
    Ug:=Append(Ug,ug);
end for;

// compute G
G:=[x[i]- qq[i] : i in [1..t]] cat Ug;

// compute A1 and A2 random affine linear bijections
L1:=Random(GLm); c1:=Random(Vm);		
A1:=[&+[L1[j][i]*z[i] : i in [1..m]]+c1[j]: j in [1..m]];
L2:=Random(GLn); c2:=Random(Vn);		
A2:=[&+[L2[j][i]*x[i] : i in [1..n]]+c2[j]: j in [1..n]];

// compute Gpub
GA2:=[(Evaluate(G[i],A2)) : i in [1..m]];
Gpub:=[Evaluate(A1[i],GA2) : i in [1..m]];

// target input and corresponding output to attack
secIPT:=Random(Vn);
OPT:=[Evaluate(Gpub[i],Eltseq(secIPT)) : i in [1..m]];
\end{verbatim}

\subsection{The Code for Method 1: Gr\"obner bases algorithms}

The following code produces the output from~\S\ref{section:method1}. Theorem \ref{thm.1} implies that, if we compute the quadratic equations corresponding to $x-\q(y)$ by Gaussian elimination, then reduce $G_{\mathrm{pub}}$ modulo a Gr\"obner basis of the space of quadrics that we found, this produces the desired quadratic system. We follow this approach in the toy example, even though it results in slightly more computations than necessary, as this allows us to
take advantage of the pre-implemented functions in Magma~\cite{BCP97}.

\begin{verbatim}
Gpubeval:=[Gpub[i]-OPT[i] : i in [1..m]];

// recover the quadratic equations in Gpubeval by solving a linear system
Mon34:=[x[i1]*x[i2]*x[i3]*x[i4] : 
            i1,i2,i3,i4 in [1..n] | i1 le i2 and i2 le i3 and i3 le i4] 
        cat [x[i1]*x[i2]*x[i3] : i1,i2,i3 in [1..n] | i1 le i2 and i2 le i3];
Acf:=[];
for f in Gpubeval do
    cf:=[MonomialCoefficient(f,mon34) : mon34 in Mon34];
    Acf:=Append(Acf,cf);
end for;
AcfM:=Matrix(Acf);
sl,K:=Solution(AcfM,VectorSpace(Fq,#cf)!0);

// Gp2 contains the quadratic equations
// Gp4 contains the other equations (cubic and quartic)
Gp2:=[]; Gp4:=[];
B1:=Basis(K); 
BB:=ExtendBasis(B1, Vm); B2:=[BB[i] : i in [#B1+1 .. m]];
for bb in B1 do
    Gp2:=Append(Gp2, &+[bb[j]*Gpubeval[j] : j in [1..m]]); end for;
for bb in B2 do
    Gp4:=Append(Gp4, &+[bb[j]*Gpubeval[j] : j in [1..m]]); 
end for;

I2:=Ideal(Gp2);
GB2:=GroebnerBasis(I2);
BR:=[b: b in GB2 | TotalDegree(b) le 4];

// Ured contains the reduced equations (w.r.t. the quadratic equations)
Ured:=[];
for f in Gp4 do
    g:=NormalForm(f,GB2);
    Ured:=Append(Ured,g);
end for;

// Gred1 is the reduced quadratic system
Gred1:= Gp2 cat Ured;
\end{verbatim}

\subsection{The Code for Method 2: Higher Order Linearization Equations (HOLE)}

\begin{verbatim}
tot:=(n+m+1)*(n+m+2) div 2; // lower bound on the number of evalutaions to perform

// recover the quadratic relations between input and output variables
E:=[];
for count in [1..tot+20] do
    ipt:=Random(Vn);
    opt:=[Evaluate(Gpub[i],Eltseq(ipt)) : i in [1..m]];
    cf:=[opt[i1]*opt[i2] : i1,i2 in [1..m] | i1 le i2] 
        cat [opt[i1]*ipt[i2] : i1 in [1..m], i2 in [1..n]]
        cat [ipt[i1]*ipt[i2] : i1,i2 in [1..n] | i1 le i2] 
        cat opt cat Eltseq(ipt) cat [1];
    E:=Append(E,cf); 
end for;

VV:=VectorSpace(Fq,#E);
EM:=Transpose(Matrix(E));
sol,K:=Solution(EM, VV!0);
BE:=Basis(K);

// given the target output, construct the equations for the target input
// Gred2 is the reduced quadratic system
Gred2:=[];
for bb in BE do
    cf:=[OPT[i1]*OPT[i2] : i1,i2 in [1..m] | i1 le i2]
        cat [OPT[i1]*x[i2] : i1 in [1..m], i2 in [1..n]] 
        cat [x[i1]*x[i2] : i1,i2 in [1..n] | i1 le i2] cat OPT cat x cat [1];
    newf:=&+[bb[i]*cf[i] : i in [1..#cf]];
    Gred2:=Append(Gred2,newf); 
end for;
\end{verbatim}

\end{appendix}

	

\bibliographystyle{siam}
\bibliography{biblio}

\begin{thebibliography}{10}

\bibitem{BCP97}
{\sc W.~Bosma, J.~Cannon, and C.~Playoust}, {\em The {Magma} algebra system
  {I}: The user language}, Journal of Symbolic Computation, 24 (1997),
  pp.~235--265.

\bibitem{CCV24}
{\sc M.~Calderini, A.~Caminata, and I.~Villa}, {\em A new multivariate
  primitive from {CCZ} equivalence}, Journal of Cryptology, 38 (2025).

\bibitem{CG21}
{\sc A.~Caminata and E.~Gorla}, {\em Solving multivariate polynomial systems
  and an invariant from commutative algebra}, in Arithmetic of Finite Fields,
  vol.~12542 of Lecture Notes in Computer Science, Springer, Cham, 2021,
  pp.~3--36.

\bibitem{CG23}
\leavevmode\vrule height 2pt depth -1.6pt width 23pt, {\em Solving degree, last
  fall degree, and related invariants}, Journal of Symbolic Computation, 114
  (2023), pp.~322--335.

\bibitem{CP19}
{\sc A.~Canteaut and L.~Perrin}, {\em On {CCZ}-equivalence, extended-affine
  equivalence, and function twisting}, Finite Fields and their Applications, 56
  (2019), pp.~209--246.

\bibitem{CCZ}
{\sc C.~Carlet, P.~Charpin, and V.~Zinoviev}, {\em Codes, bent functions and
  permutations suitable for {DES}-like cryptosystems}, Designs, Codes and
  Cryptography, 15 (1998), pp.~125--156.

\bibitem{Ding2020}
{\sc J.~Ding, A.~Petzoldt, and D.~S. Schmidt}, {\em Oil and Vinegar}, Springer
  US, New York, NY, 2020, pp.~89--151.

\bibitem{GMP22}
{\sc E.~Gorla, D.~Mueller, and C.~Petit}, {\em Stronger bounds on the cost of
  computing {G}röbner bases for {HFE} systems}, Journal of Symbolic
  Computation, 109 (2022), p.~386 – 398.

\bibitem{Huang2018207}
{\sc M.-D.~A. Huang, M.~Kosters, Y.~Yang, and S.~L. Yeo}, {\em On the last fall
  degree of zero-dimensional weil descent systems}, Journal of Symbolic
  Computation, 87 (2018), p.~207 – 226.

\bibitem{KS98}
{\sc A.~Kipnis and A.~Shamir}, {\em Cryptanalysis of the oil and vinegar
  signature scheme}, in Annual international cryptology conference, Springer,
  1998, pp.~257--266.

\bibitem{MI}
{\sc T.~Matsumoto and H.~Imai}, {\em Public quadratic polynomial-tuples for
  efficient signature-verification and message-encryption}, in Advances in
  Cryptology—EUROCRYPT’88: Workshop on the Theory and Application of
  Cryptographic Techniques Davos, Switzerland, May 25--27, 1988 Proceedings 7,
  Springer, 1988, pp.~419--453.

\bibitem{Pat95}
{\sc J.~Patarin}, {\em Cryptanalysis of the {Matsumoto} and {Imai} public key
  scheme of {E}urocrypt '88}, in Advances in Cryptology -- CRYPTO '95, vol.~963
  of Lecture Notes in Computer Science, Springer, Berlin, 1995, pp.~248--261.

\bibitem{Sal23}
{\sc F.~Salizzoni}, {\em An upper bound for the solving degree in terms of the
  degree of regularity}, preprint, arXiv:2304.13485.

\end{thebibliography}

\end{document}